\documentclass[12pt,aps,prc,amssymb,floatfix]{revtex4}
\usepackage{graphicx}
\usepackage{epsfig}    

\def\lsim{\mathrel{\rlap{
\lower4pt\hbox{\hskip-3pt$\sim$}}
    \raise1pt\hbox{$<$}}}     
\def\gsim{\mathrel{\rlap{
\lower4pt\hbox{\hskip-3pt$\sim$}}
    \raise1pt\hbox{$>$}}}     

\begin{document}



\begin{center}
{\large {\bf TOWARDS SEARCHING FOR A MIXED PHASE OF STRONGLY
INTERACTING QCD MATTER  AT THE JINR NUCLOTRON}}

 \vspace*{5mm}

A.N.~Sissakian$^a$, A.S.~Sorin$^{a}$, M.K.~Suleymanov$^b$,
V.D.~Toneev$^a$, and G.M.~Zinovjev$^c$ \\
\vspace*{3mm} {\it
$a)$ BLTP JINR, 141980 Dubna, Moscow region, Russia\\
$b)$ VBLHE JINR, 141980 Dubna, Moscow region, Russia\\
$c)$ BITP NAS, Kiev, Ukraine } \\
\end{center}

\vspace*{5mm}

{\small { \centerline{\bf Abstract} Available experimental data in
respect to a possible formation of a  strongly interacting mixed
phase  are outlined. A physical program is formulated for new
facilities being opend in Dubna for
acceleration of heavy ions with an energy up to  5 AGeV.}}\\[5mm]

\vspace*{3mm}

Over the last 25 years a lot of efforts have been made to search
for new states of strongly interacting matter under extreme
conditions of high temperature and/or baryon density, as predicted
by Quantum Chromodynamics (QCD). These states are relevant to
understanding the evolution of the early Universe after Big Bang,
the formation of neutron stars, and the physics of heavy-ion
collisions. The latter is of great importance since it opens a way
to reproduce these extreme conditions in the Earth laboratory.
This explains a permanent trend of leading world research centers
to construct new heavy ion accelerators for even higher colliding
energy.

\begin{figure}[h]
 \includegraphics[width=9cm]{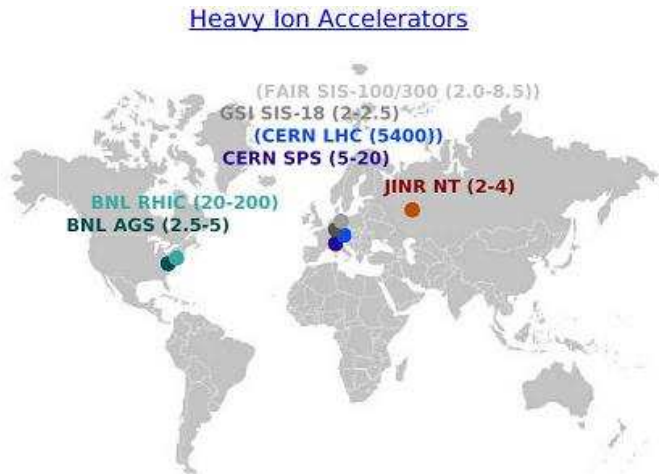}
  \caption{  The present and future accelerators of relativistic nuclear beams. For each
facility the name of the laboratory, the name of the accelerator
and the energy range (center-of-mass energy per nucleon-nucleon
pair $\sqrt{s_{NN}}$ in GeV) are given~\cite{Gprog}. }
  \label{figj1}
\end{figure}

Looking  at the list of heavy-ion accelerators one can see that
after the first experiments at the Dubna Synchrophasotron,
heavy-ion physics successfully developed at Bevalac (Berkley) with
the bombarding energy to $E_{lab} \sim 2$ AGeV, AGS (Brookhaven)
$E_{lab} \sim 11$ AGeV, and SPS (CERN) $E_{lab} \sim 160$ AGeV.
The first two machines are closed now. The nuclear physical
programs at SPS as well as at SIS (GSI, Darmstadt, $E_{lab} \sim
1$ AGeV) are practically completed. The new relativistic heavy-ion
 collider (RHIC, Brookhaven) is intensively working in the
ultrarelativistic energy range $\sqrt{s_{NN}}\sim 200$ GeV to
searching for signals of the quark-gluon plasma formation. In this
respect, many hopes are related to the Large Hadron Collider (LHC,
CERN) which will start to operate in  the TeV region in two-three
years. The low-energy scanning program at SPS (NA49 Collaboration)
revealed an interesting structure in the energy dependence of some
observables at $E_{lab} \sim 20-30$ AGeV, which can be associated
with the exit of an excited system into a decofinement state. This
fact essentially stimulated a new large project FAIR GSI
(Darmstadt) for studying compressed baryonic matter in a large
energy range of $E_{lab} =10-30$ AGeV which should come into
operation after 2013 year~\cite{GSI300}. The general scheme how
available accelerators are spread over the world is presented in
Fig.\ref{figj1}

On the other hand, in JINR there is a modern superconducting
accelerator, Nuclotron, which has not realized its planned
parameters yet. The Veksler and Baldin Laboratory of High Energy
has certain experimental facilities and large experience to work
with heavy ions. This study may actively be supported by
theoretical investigations of the Bogoliubov Laboratory of
Theoretical Physics. The paper aims to give an overview of
experimental results on manifestation of some peculiarities of
strongly interacting QCD matter, observed in relativistic heavy
ion collisions, and to present arguments that acceleration of
heavy ions like $Au$ at the Nuclotron up to the maximal planned
energy $E_{lab}=5$ AGeV, allows one to study properties of hot and
dense QCD matter to be competitive at the world level.

\begin{figure}
 \includegraphics[width=8cm]{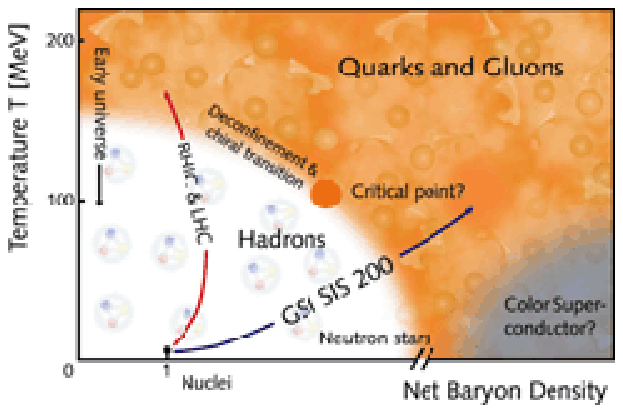}
  \caption{Artist's view of the phase diagram~\cite{GSI300} }
  \label{fig1j}
\end{figure}

A convenient way to present a variety of possible states of
strongly interacting matter is a phase diagram in terms of
temperature $T$ and baryon chemical potential $\mu_B$ (or baryon
density $n_B$), as shown in Fig.\ref{fig1j}. This schematic
picture shows in which region of the diagram the given phase is
realized and which colliding energies are needed to populate this
region. In particular, the boundary of the deconfinement and/or
chiral symmetry restoration  transition may be reached even below
bombarding energies planned in the FAIR GSI project but the
nuclear matter compression should be high enough.

\begin{figure}[h]
  \includegraphics[width=7.3cm]{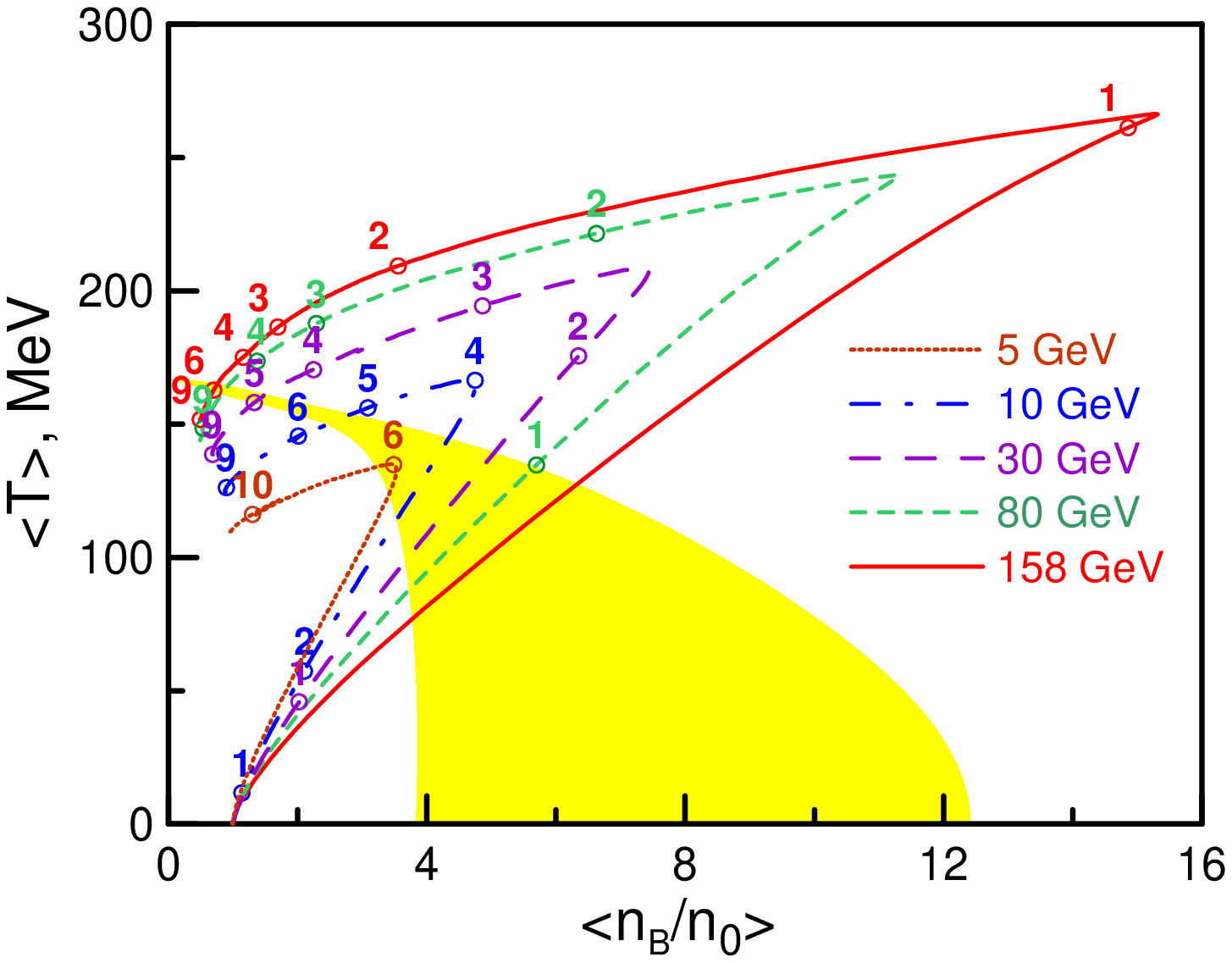}
  \includegraphics[width=7.3cm]{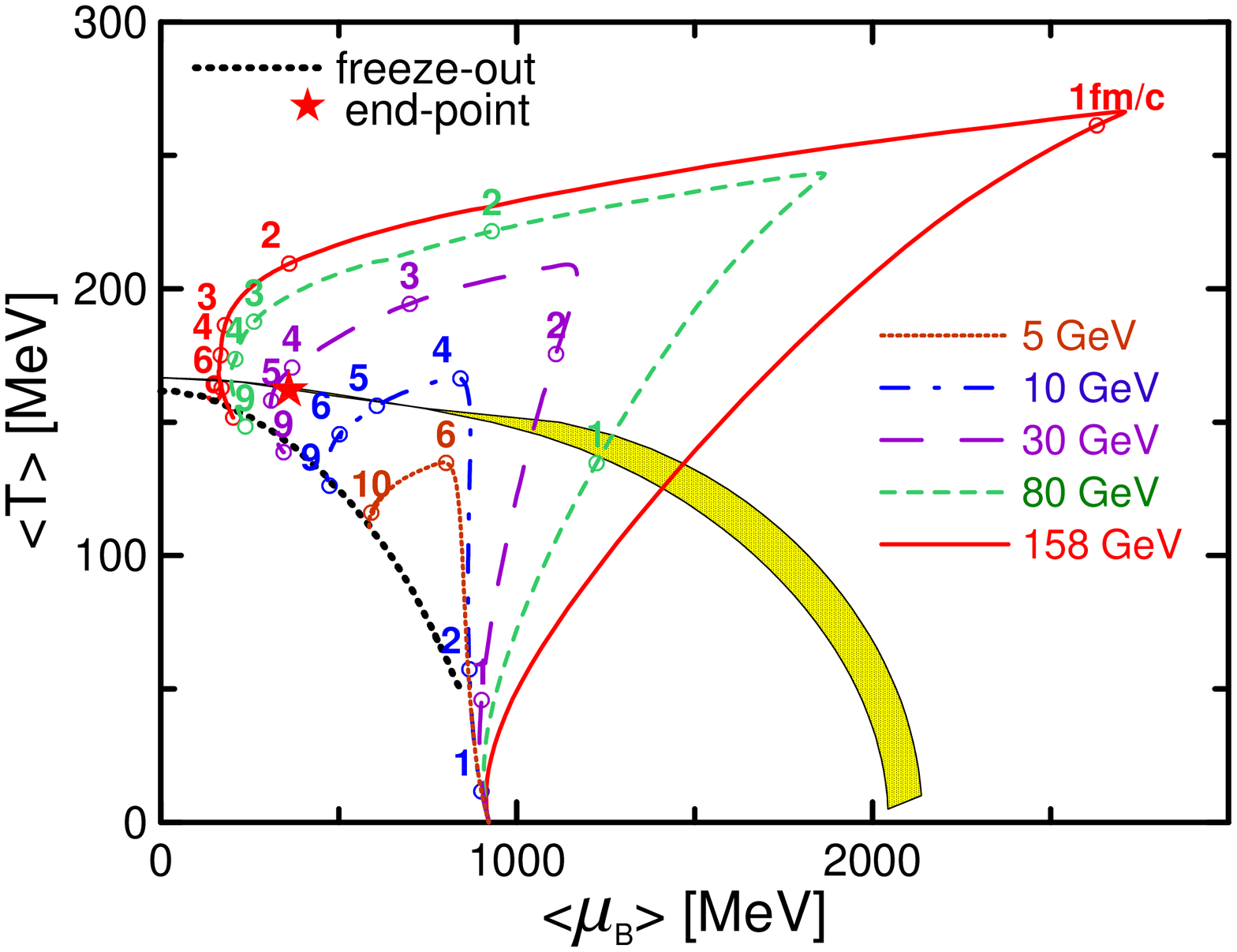}
  \caption{Dynamical trajectories for central ($b=2fm$) $Au+Au$ collisions in $T-n_B$
  (left panel)
  and  $T-\mu_B$ (right panel) plane for various bombarding energies calculated within
  the relativistic 3-fluid hydrodynamics~\cite{IRT05}. Numbers
  near the trajectories are the evolution time moment.  Phase boundaries are
  estimated in a two-phase bag model.
  }
  \label{fig2-3j}
\end{figure}

This point is illustrated in more detail in Fig.\ref{fig2-3j}.

As is seen, a system, formed in a high energy collision, is fast
heated and compressed and then starts to expand slowly reaching
the freeze-out point which defines observable hadron quantities.
Indeed, at the maximal achievable Nuclotron energy $E_{lab}=5$
AGeV the system "looks" into the mixed phase  for a short time
(the left part of Fig.\ref{fig2-3j}). However, uncertainties of
these calculations are still large. If heavy masses for $u,d$
quarks are assumed, the phase boundary is shifted towards higher
$\mu_B$ (the right part in Fig.\ref{fig2-3j}). On the other hand,
these dynamical trajectories were calculated for a pure hadronic
gas equation of state, and the presence of a phase transition may
noticeably change them. In addition, near the phase transition the
strongly interacting QCD system behaves like a liquid rather than
a gas, as was clarified recently at small $\mu_B$ from both
quark~\cite{ZS} and hadronic~\cite{V04} side. As to high $\mu_B$
values, it is a completely open question. One should note also
that as follows from lattice QCD calculations for $\mu_B \approx
0$ the deconfinement temperature practically coincides with the
transition temperature for chiral restoration symmetry while for
baryon rich matter it is still an open question.

\begin{figure}[h]
  \includegraphics[width=9cm]{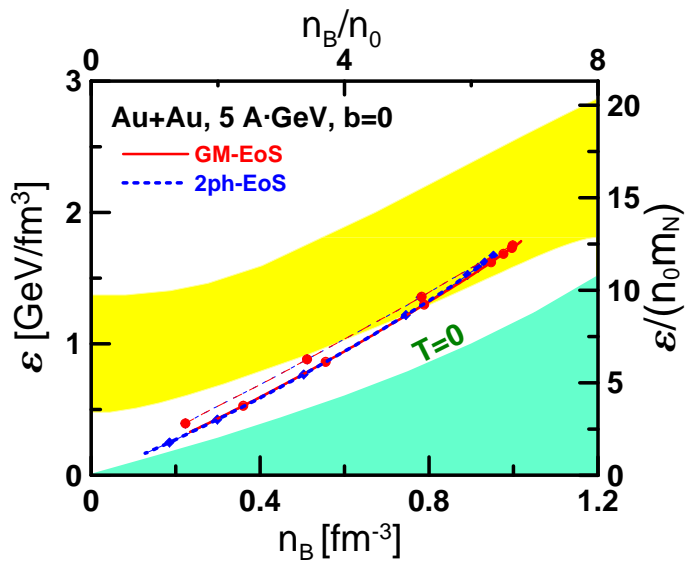}
  \caption{Dynamical trajectory in the $\varepsilon -n_B$-plane for
  central Au+Au collisions calculated with
  two equations of state: pure hadronic (solid line) and with first
  order phase transition (dashed). Spatial averaging is done over
  the cube with the 4 $fm$ sides and Lorentz contracted in the longitudinal
  direction~\cite{I05}. The shaded regions
  correspond to the mixed phase (upper one) and the non-reachable domain with
  the boundary condition $T=0$, respectively.
  }
  \label{fign_eps}
\end{figure}

In Fig.\ref{fign_eps}, dynamical trajectories in the
$\varepsilon-n_B$ plane for the top Nuclotron energy are given for
two equations of state, without and with the first order phase
transition~\cite{IRT05,I05}. The shaded band corresponds to the
mixed phase of more restrictive EoS with heavy quark masses, as
shown in the right panel of Fig.\ref{fig2-3j}. Nevertheless, both
trajectories, being close to each other, spend some time in the
mixed phase. The main difference between the results presented in
Fig.\ref{fig2-3j} and Fig.\ref{fign_eps} comes from the different
averaging procedures used: in the first case, thermodynamic
quantities were averaged over the whole volume of the interacting
system, while in the second case, it was carried out only over a
cube of 4 fm sides placed at the origin and Lorentz contracted
along the colliding axis. Thus, for central collisions at the 5
AGeV Nuclotron energy even if an average state of the whole
strongly interacting system does not approach the mixed phase, an
essential part of the system volume will have a certain time in
this mixed phase. An experimental consequence  is that an expected
observable signal for reaching the mixed phase should be rather
weak.

\begin{figure}[h]
\hspace{15mm}\includegraphics[width=8.5cm]{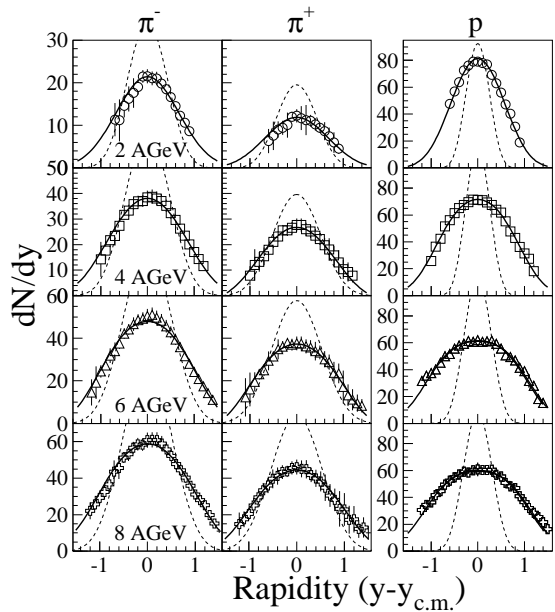}
\caption{The $\pi^+ ,\pi^-$ and proton rapidity distributions in
central Au+Au collisions at bombarding energies 2, 4, 6, and 8
AGeV \cite{Klay03}. Solid lines are drawn along experimental
points. For comparison, Gaussian distributions are given by dashed
curves.}
 \label{figj18}
\end{figure}

\begin{figure}[h]
\includegraphics[width=8cm,angle=-90]{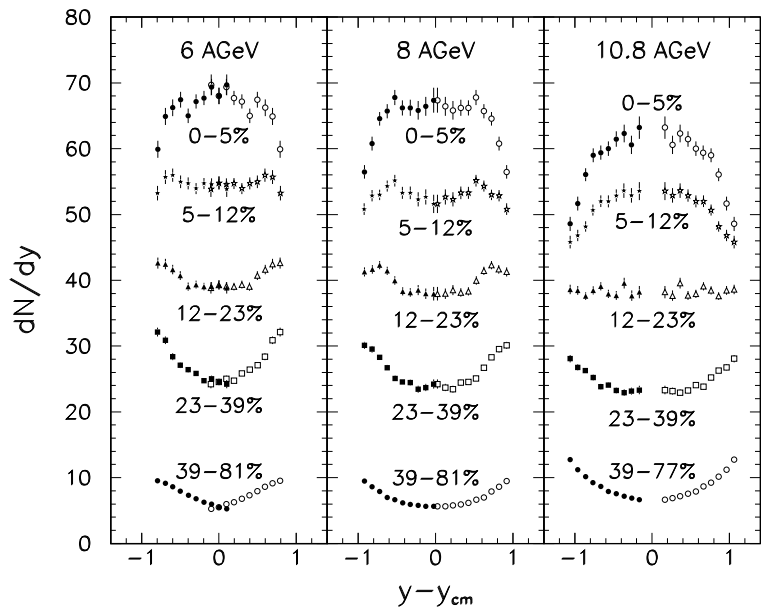}
\caption{Proton rapidity distributions for different centrality of
Au+Au collisions at 6, 8, and 10.8 AGeV   \cite{Holzman02}. Filled
symbols are measured points, empty ones are their reflection with
respect to the $y=y_{cm}$ line. Numbers under distributions give
the centrality percentage.} \label{figj19}
\end{figure}

It is hard to believe that some irregular structure like that at
$E_{lab} \sim 30$ AGeV ~\cite{Gazd05} can manifest itself at the
Nuclotron energy. So the global observables  are expected to be
quite smooth with energy. Indeed, as is seen from the results
presented in Figs.\ref{figj18} and \ref{figj19} covering the
region of $\sim$ 2-10 AGeV. The shape of rapidity spectra for
newly produced mesons is bell-like but cannot be described by a
single Gaussian due to flow effects. Note that a number of $\pi^+$
is not equal to that of $\pi^-$ though their difference
essentially decreases with the bombarding energy. So the isotopic
degree of freedom should properly be taken into account under
theoretical consideration in the Nuclotron energy range. As to
proton spectra, their shape is yet wider, ranging from the target
to projectile rapidity. At $E_{lab}\sim $ 10 AGeV it is getting
flat at the middle rapidity and may be described by a
superposition of two Gaussian to SPS energies~\cite{IRT05} and of
three Gaussian at the RHIC energies. The shape of baryon rapidity
distributions strongly varies with the impact parameter (see
Fig.\ref{figj19}) taking U-shape for peripheral collisions. The
noted trends are continued with  increase in  energy and governed
mainly by the stopping power of colliding matter. To see any
effect of phase transformation in a global rapidity distribution,
a phase like that should be created in a large volume and should
live for a long time. Possibly, it is seen as appearance of the
third midrapidity source at the RHIC energies~\cite{Wol05}.

 Thus, sensitivity of global observables to possible phase
 transitions is expected to be weak, but it might not  be the case for more delicate
characteristics. In any case, due to the proximity of the phase
diagram region under discussion to the confinement transition and
chiral symmetry restoration, some precursory phenomena cannot be
excluded at a bombarding energy of about 5 AGeV, which opens a new
perspective for physical investigations at the Dubna Nuclotron.

Below some arguments in favor of this statement are given:

\begin{itemize}
\item  Properties of hadrons  are expected to change in hot and/or
dense baryon matter~\cite{TK94,BR96}.  This change concerns
hadronic masses and widths, first of all for the $\sigma$-meson as
the  chiral partner of pions which characterizes a degree of
chiral symmetry violation and can serve as a "signal" of its
restoration as well as the mixed phase formation. Rare decays in
matter of vector mesons (particularly $\rho $ and $\omega$) are
also very attractive.
\begin{figure}[h]
  \includegraphics[width=7cm]{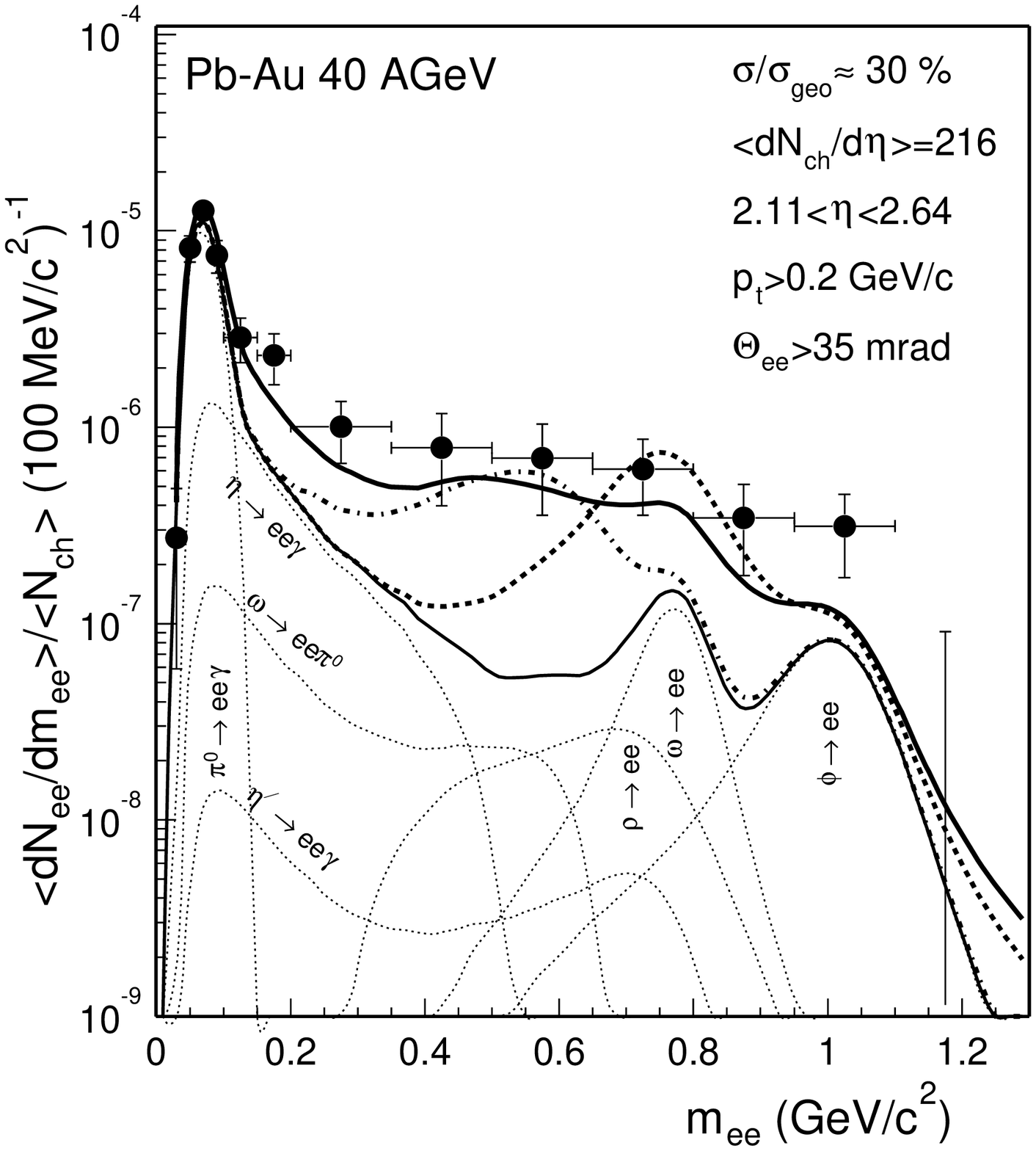}
 \includegraphics[width=7.7cm]{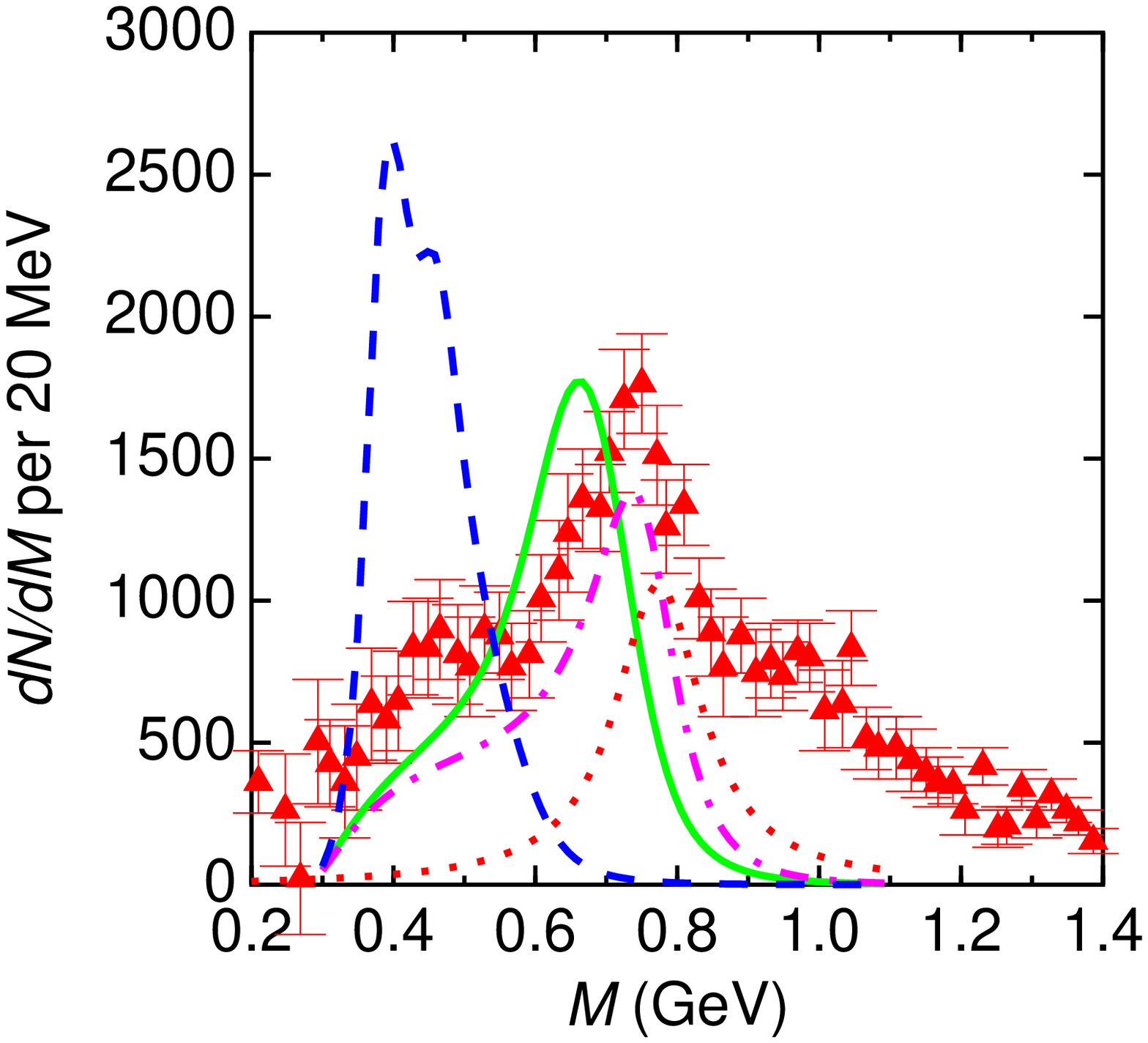}
  \caption{Left panel: $e^+e^-$ invariant spectra from central $Pb+Au$ (40 GeV)
  a collisions~\cite{CERES40}. Thin solid and dotted lines are hadronic
  cocktail and the calculated results for free $\rho$ mesons, respectively.
  Appropriate thick solid and dash-dotted lines are calculated in the
  Rapp-Wambach~\cite{RW00} and Brown-Rho~\cite{BR96} scenarios.
  Contributions of different channels are shown as well. \\
Right panel: Invariant mass distribution of dimuons from
  semi-central In+In collisions
  at the beam energy 158 $A$GeV. Experimental points are from~\cite{NA60-QM05}.
  Solid and dashed curves are calculated~\cite{ST05} in the dropping
  mass scenario
  using the $\rho$-mass modification factors as density
  and temperature-density  dependent, respectively. Dash-dotted curve neglects
  any in-medium modification.
  Dotted line indicates the hydrodynamically
  calculated $\rho$-meson decay at the freeze-out.}
  \label{fig4j-Na60}
\end{figure}

The presence of in-medium modification of $\rho$-mesons has been
proved in the CERES experiments, see an example in
Fig.\ref{fig4j-Na60}. The observed essential enhancement of
low-mass ($0.2 \lsim M \lsim 0.7$) lepton pairs, as compared to
free hadron decays, is due to the influence of hot and dense
nuclear matter on properties of the $\rho$-meson spectral
function. Both the Brown-Rho  scaling hypothesis~\cite{BR96}
assuming a dropping $\rho$ mass and a strong broadening, as found
in the many-body approach by Rapp and Wambach~\cite{RW00}, result
in reasonable agreement with experiment. However, a poor
resolution in the dielectron mass prevents the discrimination of
different physical scenarios of this effect in the CERES
experiments. The situation is noticeably better in the recent NA60
experiment with muon pairs~\cite{NA60-QM05}. As is seen from the
right panel of Fig.\ref{fig4j-Na60}, the quality of dimuon data is
much higher allowing a good resolution of the spectral function.
The depicted comparison with the dropping mass scenario shows a
strong influence of the final result on the hypothesis used for
in-medium dependence of the mass modification factor~\cite{ST05}.
However, it seems to be impossible to describe simultaneously
dielectron and dimuon experimental data within the dropping mass
scenario under the same assumptions. Similar data are absent for
heavy ions in the Nuclotron energy range.

As to in-medium $\sigma$-meson decay, some indications were
obtained in reactions induced by pions and
photons~\cite{CHAOS,CBall,TAPS2}. In Fig.\ref{figCHAOS}, the
relative pion pair abundance $C_{\pi\pi}^A$, defined as
$$C_{\pi\pi}^A =\frac {
\sigma^A (M_{\pi\pi})}{ \sigma_T^A} \ / \ \frac {
\sigma^N(M_{\pi\pi})}{ \sigma_T^N}~,$$ is presented for different
isotopic states of observed pions versus the invariant pion pair
mass $M_{\pi\pi}$.
\begin{figure}[h]
 \hspace*{2cm} \includegraphics[width=12cm]{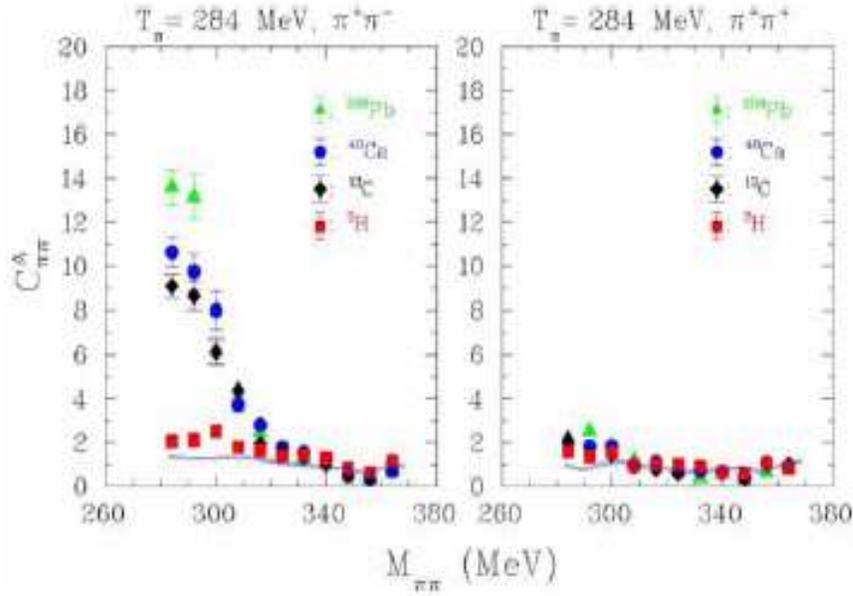}
  \caption{Invariant mass distribution of pion pairs from $\pi$-nucleus
  interactions~\cite{CHAOS}. }
  \label{figCHAOS}
\end{figure}
A sizable enhancement is observed at low $M_{\pi\pi}$ in the case
of $\pi^+ \to \pi^-\pi^-$ reaction but not for the $\pi^+ \to
\pi^+\pi^+$ case~\cite{CHAOS}. This enhancement may be related to
a possible $\pi\pi$ scattering in a nucleus via formation of a
virtual scalar $\sigma$ meson which is forbidden for the
$\pi^+\pi^+$ state. The effect is getting stronger for heavier
nuclei. Recently, using the ELSA tagged facilities in Bonn the
in-medium modification of the $\omega$-meson has first been
observed in the reaction $\gamma +A \to \omega +X \to \pi^0+\gamma
+X$~\cite{TAPS2}. A small shift in the $\omega$ mass was detected.
Note, however, that in $\pi$- and $\gamma$-induced reactions we
deal with comparatively low baryon density states, $n_ B\sim
(1-2)n_0$. There are no similar experiments with heavy ions.

Nevertheless, there are theoretical proposals to probe chiral
symmetry restoration in the vicinity of the phase transition
boundary. In particular, it was shown~\cite{CH98,VKBRS98} that a
two-photon decay of the $\sigma$-meson formed as an intermediate
state in $\pi\pi$ scattering may be a very attractive signal. As
depicted in Fig.\ref{fig6j}, at temperature in the vicinity of
\begin{figure}[h]
 \hspace*{3.cm} \includegraphics[width=7.5cm,angle=-90]{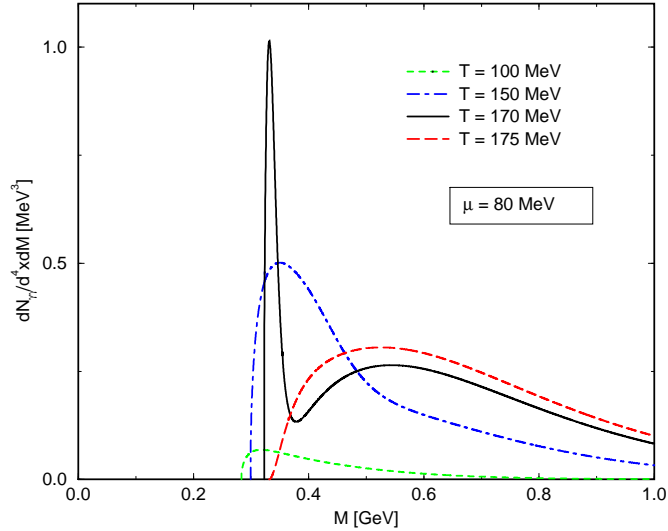}
  \caption{Invariant mass spectra of $2\gamma$ at $\mu_B=80$ MeV
  and different temperatures~\cite{VKBRS98}. }
  \label{fig6j}
\end{figure}
 the phase transition, when $m_\sigma \sim 2m_\pi$, there is an anomalous peak
in invariant mass spectra of $\gamma$ pairs which may serve as a
signal of the phase transition and formation of a mixed phase.
Certainly, there is a huge combinatorial background due to $\pi^0
\to \gamma\gamma$ decays, but the Nuclotron energy is expected to
have some advantage against higher energy accelerators because the
contribution of deconfined quarks-gluons will be negligible.

This effect may be observed in the $e^+e^-$ decay channel as
well~\cite{Weld92}.

 \item Electromagnetic probes discussed above  carry out
 information concerning the whole evolution of colliding nuclei and states which
 are realized. The bulk of produced hadrons is related to
 the freeze-out point where information on the interaction dynamics
 has essentially been erased. So the expected behavior of global
 hadronic observables is smooth. However, some peculiarities of delicate
 hadron characteristics may be found and their hints are available
 even now.

\begin{figure}[h]
 \includegraphics[width=9cm]{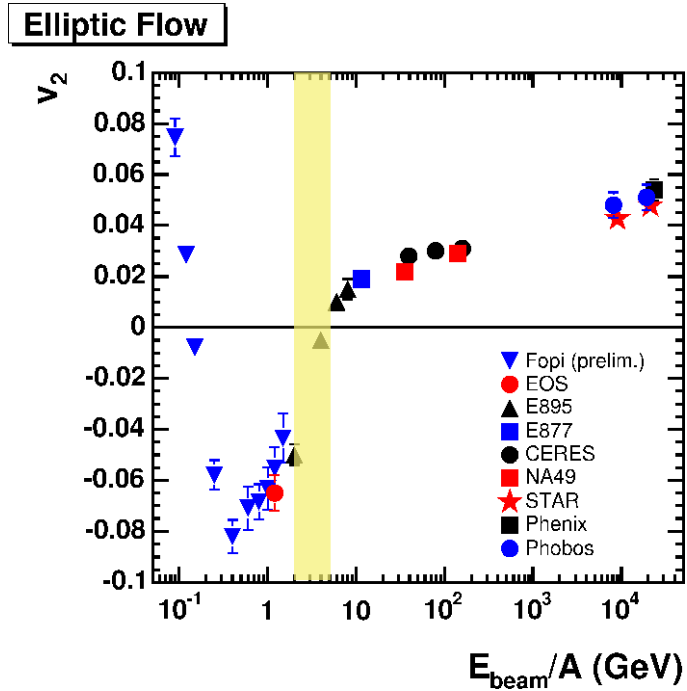}
  \caption{Excitation function of the proton elliptic flow ~\cite{St04}. The shaded
  band corresponds to the Nucltron energy.}
  \label{fig7j}
\end{figure}

 In Fig.\ref{fig7j}, the beam energy dependence of the elliptic flow coefficient $v_2$
  is presented for protons in the midrapidity  range from noncentral
  heavy-ion collisions~\cite{St04}. The elliptic
 flow characterizes the angular anisotropy  in the transverse
 momentum plane as
 $$ dN/d\phi \sim \left[ 1+2v_1 \cos (\phi)+2v_2 \cos (2\phi) \right]~, $$
 where the $\phi$ angle is measured from the reaction plane. As is
 seen,  just near the  Nuclotron energy $E_{lab}\sim 5$ AGeV the
 coefficient $v_2$ changes its sign and the transverse flow
  evolves from the out-of-plane towards in-plane emission. This fact may be treated
  as softening of the equation of state to be a precursor of the phase transition.
 The elliptic flow shows an essentially linear dependence on the
impact parameter with a negative slope  at $E_{lab}=2$ AGeV, with
a positive slope at 6 AGeV and with a near zero slope at 4
AGeV~\cite{Ch01}. This dependence serves as an important
constraint to  high-density behavior of nuclear matter for
discrimination of various equations of state.

\item Strangeness enhancement is an intriguing point of physics of
heavy ion collisions, being one of the first proposed signals of
quark-gluon plasma formation. An important experimental finding is
the observation of some structure ("horn") in the energy
dependence of reduced strangeness multiplicity at $E_{lab}\sim 30$
AGeV, predicted in~\cite{GG99} as a signal that the formed excited
system came into a deconfinement phase. As an example, in
Fig.\ref{figj7s} the $K^+/\pi^+$ average multiplicity ratio is
displayed as a function of the colliding energy. The "horn"
structure is well visible. Such a structure is absent in free
nucleon-nucleon collisions.

\begin{figure}[thb]
 \includegraphics[width=8cm]{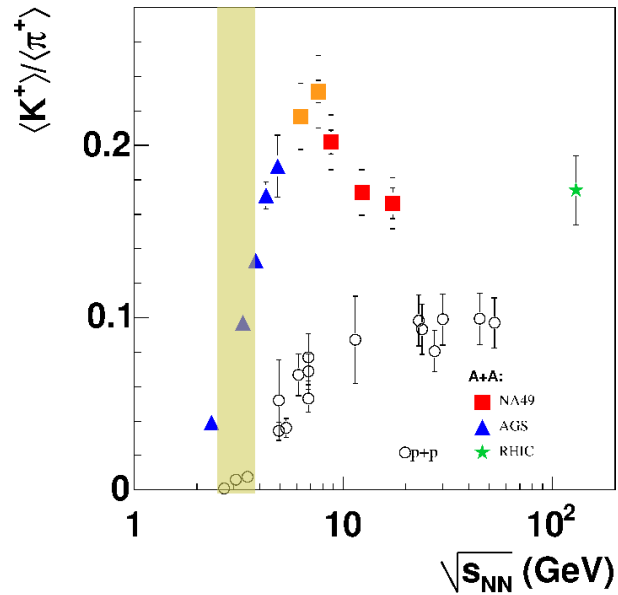}
  \caption{Colliding energy dependence of the reduced strangeness abundance of $K^+$
    mesons~\cite{Gazd04}. The shaded band corresponds to the Nuclotron energy range. }
    \label{figj7s}
\end{figure}
\begin{figure}[h]
\includegraphics[width=8cm]{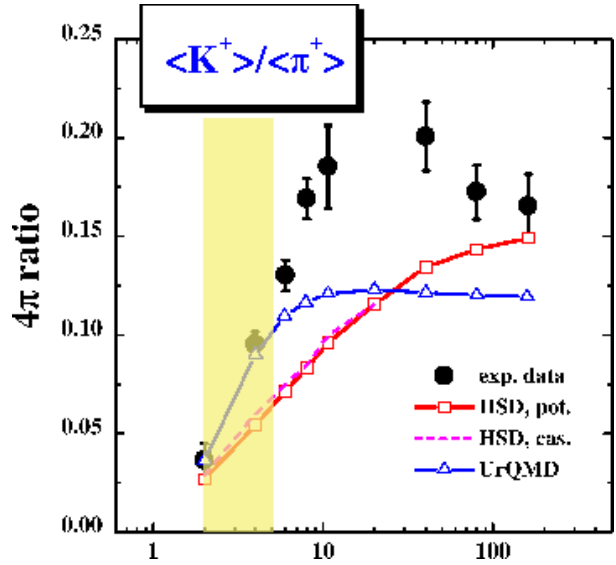}
   \caption{Beam energy dependence of the $K^+/\pi^+$ ratio.
   Curves are the results of different transport calculations~(the figure is taken
   from~\cite{WBCS02}. The shaded
   band corresponds to the Nuclotron energy range}
   \label{fig8j}
\end{figure}

The same experimental data (besides two recent points measured at
$E_{lab}=$20 and 30 AGeV) are plotted in Fig.\ref{fig8j} as a
function of the bombarding energy.  It is of great interest that
these ratios for strange hadrons are not explained by the modern
transport theory (UrQMD, HSD models). While the mean pion and kaon
multiplicities are well reproduced at the SIS and SPS energies,
the above-mentioned models essentially underestimate the $K /\pi$
ratio in the AGS energy domain. It is remarkable that the
divergence between the transport theory and experiment starts just
at the Nuclotron energy. This increases interest in future
experiments at the Nuclotron.

 \item
A strange particle ratio is not the only attractive observable.
Let us look at the Lorentz invariant quantity, the transverse
momentum distribution, presented in Fig.\ref{figj320-30}.
\begin{figure}[h]
\includegraphics[width=7.5cm,height=8.5cm]{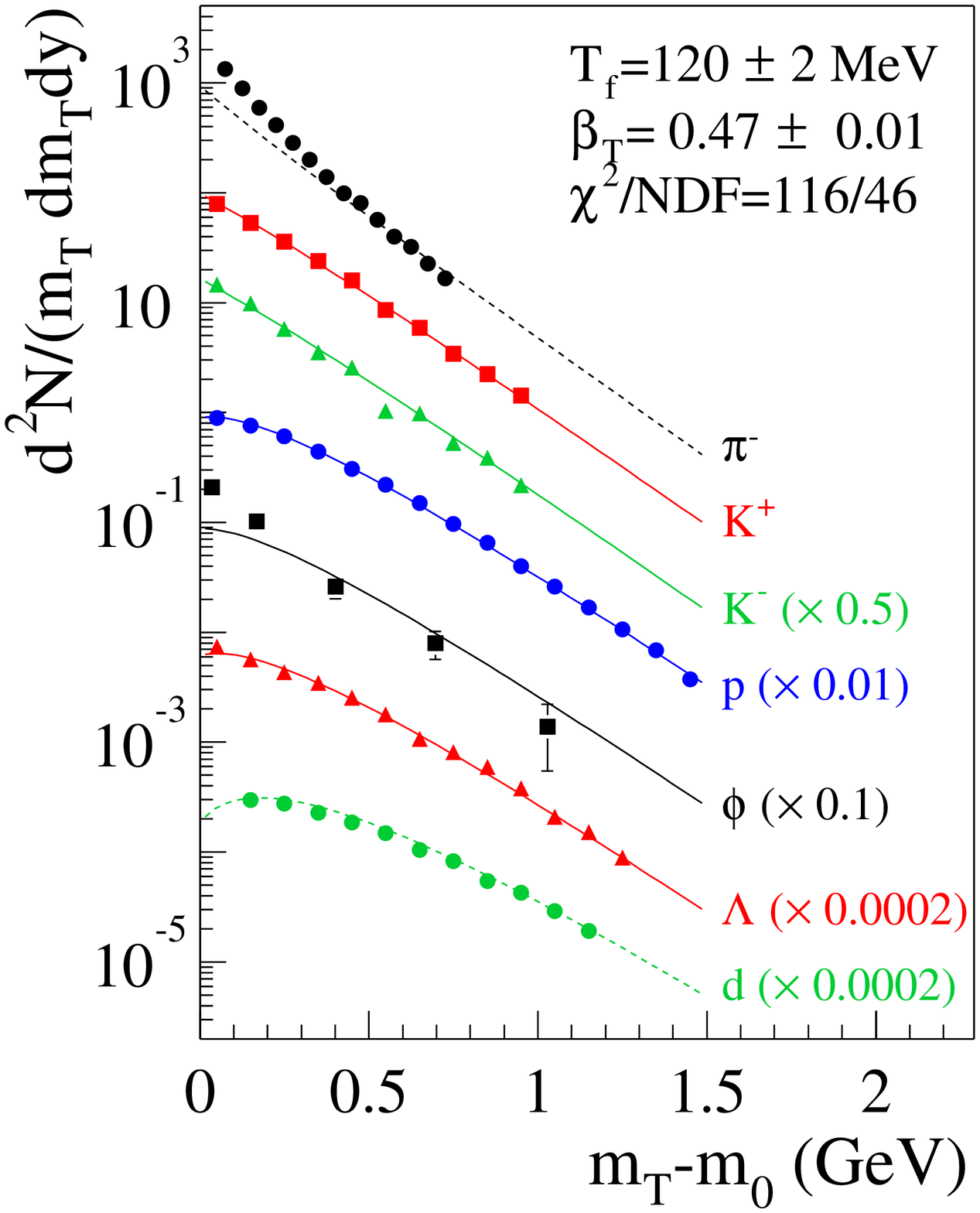}
\includegraphics[width=7.5cm,height=8.5cm]{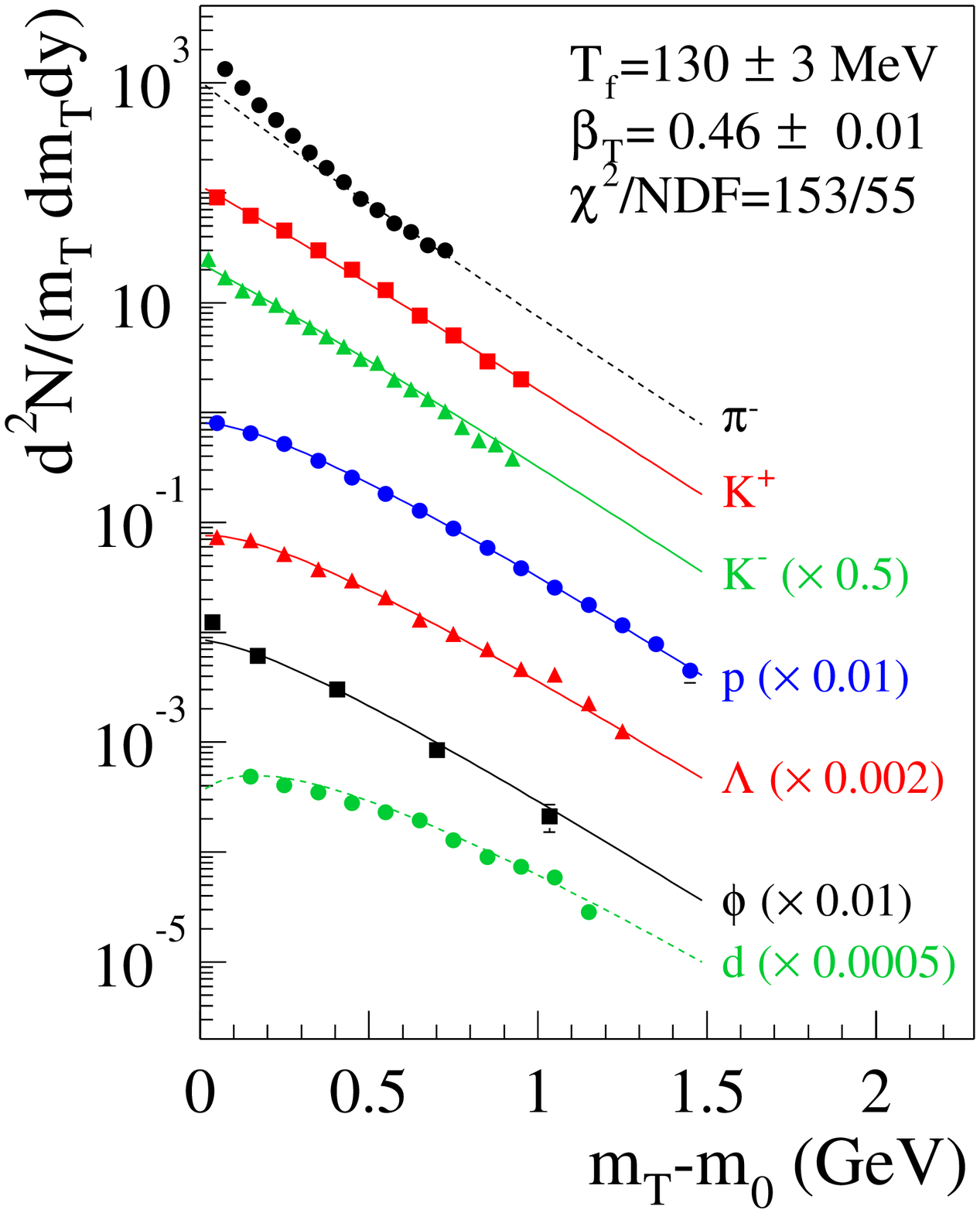}
\caption{The transverse mass spectra at midrapidity for different
hadrons produced in central Pb+Pb collisions at 20 AGeV and  30
AGeV~\cite{Gazd04}. Solid lines are results of the blast-wave
parameterization\cite{Schn94} with the parameters given in the
figure.} \label{figj320-30}
\end{figure}
\begin{figure}[h]
\includegraphics[width=12.cm]{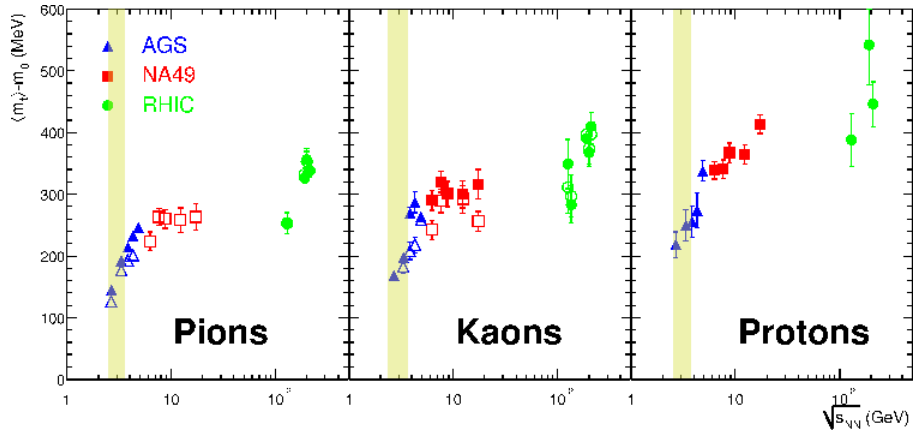} \caption{The
energy  dependence of $<m_T> - m_0$ for pions, kaons, and protons
at midrapidity for the most central Pb+Pb/Au+Au
collisions~\cite{Gazd05}. Shaded bands correspond to the Nuclotron
energy. } \label{figj11s}
\end{figure}

It is a global characteristic and no peculiarities are expected.
Indeed, experimental spectra are reasonably well described within
a simple blast-wave model~\cite{Schn94} where spectra at
freeze-out are parameterized as follows:
$${{dN}\over{dy \ m_Tdm_T}}\simeq{{m_TK_1 \ (}{{{m_T \ \cosh\rho}\over
T})} \ {I_0(}{{{p_T \ \sinh\rho}\over T})}}$$ with the boost angle
$\rho = \tanh^{-1} \beta_T$ and values of the freeze-out
temperature $T$ and transverse velocity $\beta_T$ are given in
Fig.\ref{figj320-30}. However, the behavior  of the average
transverse mass, $<m_T> - m_0$, versus  colliding energy $\sqrt
s_{NN}$, is not trivial. As follows from Fig.\ref{figj11s},  a
remarkable change in the energy dependence around
\begin{figure}[h]
\includegraphics[width=7.3cm]{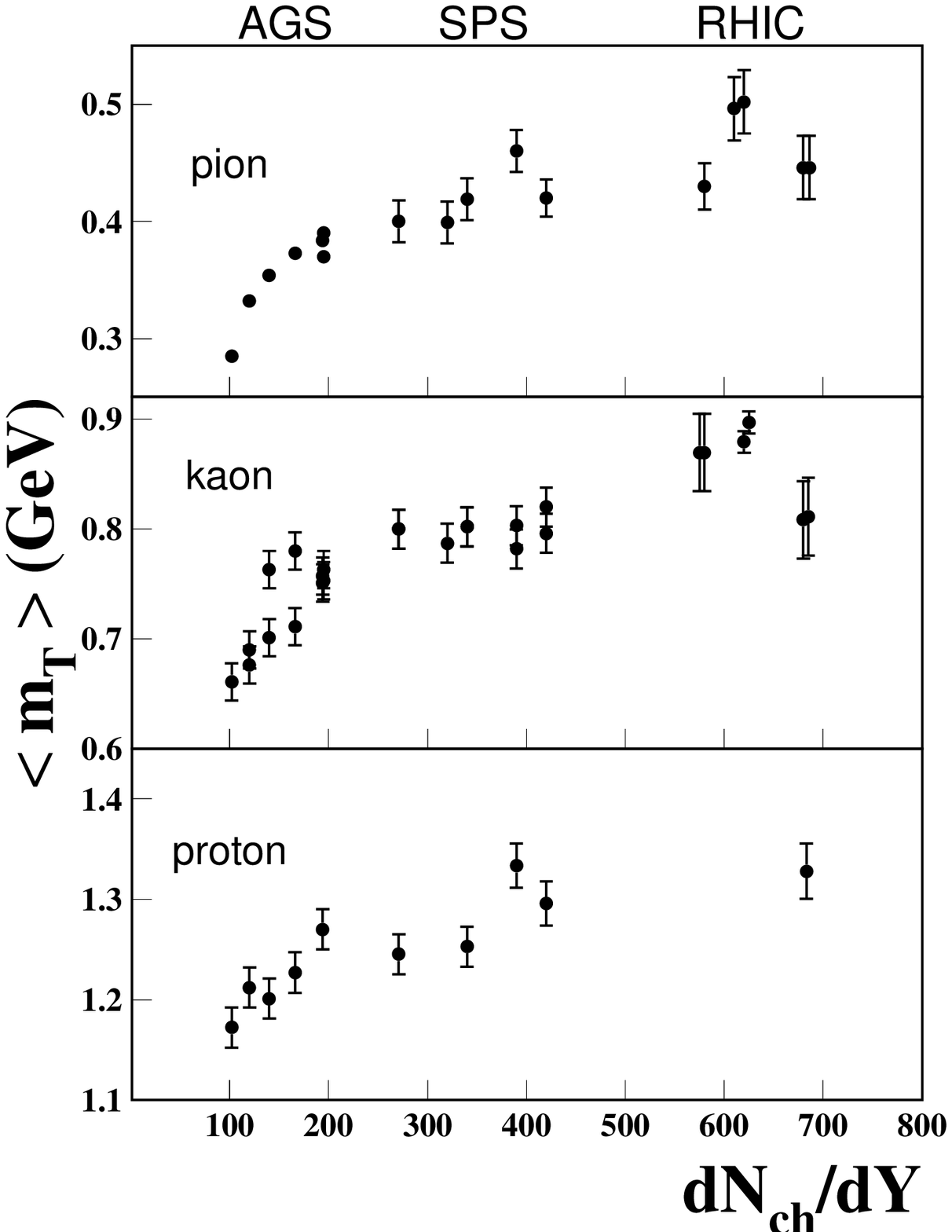} \caption{
Variation of $<m_T>$ with produced charged particles per unit
rapidity at midrapidity for central collisions corresponding to
the energy range from AGS to RHIC~\cite{MASNN03}. The error bars
reflect both the systematic and statistical errors in obtaining
$T_{eff}$.}\label{figj1dd}
\end{figure}
\begin{figure}[h]
\includegraphics[width=7.3cm]{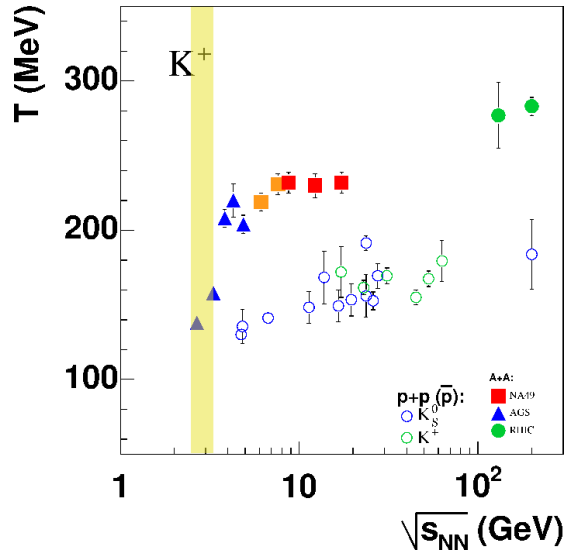} \caption{The
energy dependence of the inverse slope parameter of
kaons~\cite{Gazd05}. The shaded band corresponds to the Nuclotron
energy.}\label{figj10s}
\end{figure}
a beam energy of $\sim$30 AGeV is clearly visible for pions and
kaons exhibiting some kind of plateau. While $(<m_T> -m_0)$ rises
steeply in the AGS energy range, this rise is much weaker from low
SPS energies until RHIC energies where it starts again to rise. To
a lesser extent this change is also seen for protons. One should
emphasize that the beginning of the plateau is well correlated
with the "horn" position. Measurements at the Nucloron may specify
a pre-plateau behavior, in particular, for kaons and protons.

Among many signals of the formation of Quark Gluon Plasma (QGP),
one of the earliest is based on the relation of the
thermodynamical variables, temperature, and entropy to the average
transverse momentum and multiplicity, respectively, as was
originally proposed by Van Hove \cite{vH82}. It was argued that a
plateau in the transverse momentum beyond a certain value of
multiplicity would indicate the onset of the formation of a mixed
phase of QGP and hadrons, similar to the plateau observed in the
variation of temperature with entropy in a first order phase
transition scenario. In Fig.\ref{figj1dd}, the variation of $<m_
T>$ with charged multiplicity produced per unit rapidity $dN/dy$
around the midrapidty~\cite{MASNN03} is depicted for pions, kaons,
and protons at AGS, SPS and RHIC energies spanning the range from
2 AGeV to 200 AGeV.

\begin{figure}[h]
\includegraphics[width=8.3cm]{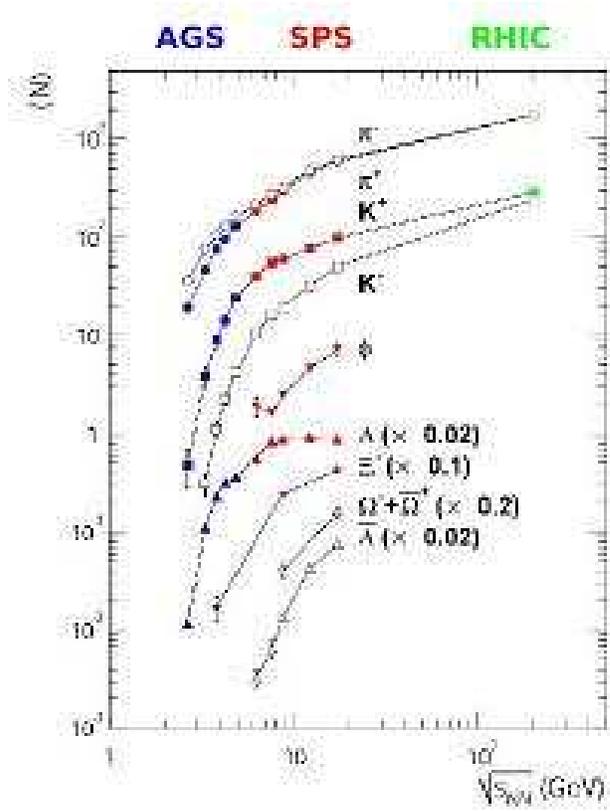}
\caption{The colliding energy dependence of the mean
multiplicities of various hadrons emitted in central Pb+Pb/Au+Au
collisions~\cite{Bl05}.}\label{figj5}
\end{figure}

  In this analysis the experimental data on transverse mass
spectra were parameterized as
\begin{equation}
\frac{dN}{m_T \ dm_T} \simeq C \exp{(- {m_T}/T_{eff}} )
\label{Teff}
\end{equation}
with the inverse slope parameter  treated as an effective
temperature $T_{eff}$. The average transverse mass of hadrons is
related with $T_{eff}$:
$${<m_T>}={T_{eff}+m+{{(T_{eff})^2}\over{m+T_{eff}}}}.~$$

From the results shown in Fig.\ref{figj1dd} one observes an
increase in $<m_T>$ with a particle density $dN_{ch}/dy$ for AGS
energies followed by a plateau for charge multiplicities
corresponding to SPS energies for all the particle types
considered: pions, kaons and protons. This may hint at the mixed
phase, possible co-existence of the quark and hadron phases. For
charge multiplicities corresponding to higher RHIC energies, the
$<m_T>$ shows an increasing trend indicating the possibility of a
pure QGP formation.

Using eq.(\ref{Teff}), one may directly  analyze data in terms of
effective temperature. As is  shown in Fig.\ref{figj10s}, the
inverse slope parameter of kaons increases in the AGS and RHIC
energy domains but it stays constant at SPS energies in natural
agreement with the particle ratio results presented above. This
feature, which is not observed in $p+p$ interactions, might be
attributed to the latent heat of a phase transition~\cite{GG99}
and it is in fact consistent with hydrodynamic model calculations
assuming a first order  phase transition~\cite{hydro04}.

\item As mentioned above, global multiplicities are expected to be
quite smooth over a large range of energy. Fig.\ref{figj5} shows
the mean multiplicities of hadrons emitted in central Pb+Pb/Au+Au
collisions as a function of energy. The results are well described
within the statistical hadron gas model~\cite{Bl05}. Note that a
number of charged pions  $(n_{\pi^+}+n_{\pi^-})\sim 200$ and kaons
$K^+\sim$ per central event at the top Nuclotron energy.
\begin{figure}[h]
 \hspace*{4cm}\includegraphics[width=10cm]{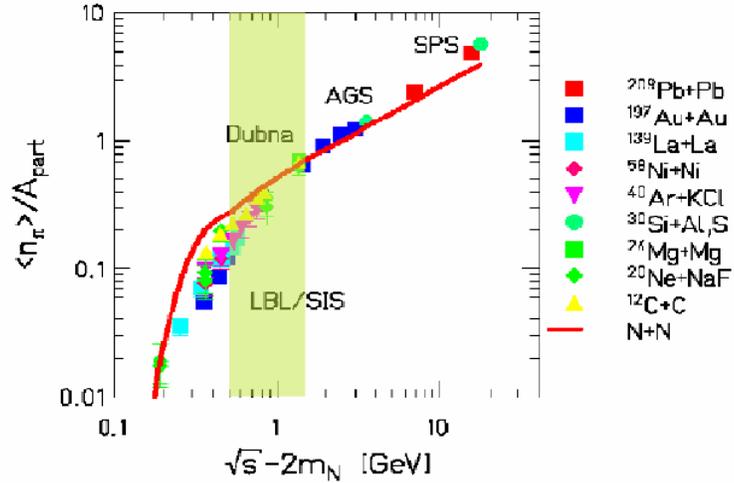}
   \caption{Pion multiplicity per participating nucleon for
   nucleus-nucleus (symbol) and nucleon-nucleon collisions (solid
   line) as a function of available energy in nucleon-nucleon collisions~\cite{SS99}.
   The shaded band corresponds to the Nuclotron energy.
    }
    \label{fig9js}
\end{figure}

However one should add that peculiarities at the  "horn" energy in
strange particle ratios and the effective temperature dependence
are also observed in  average pion number per nucleon-nucleon
interaction~\cite{Gazd05}. Such an example is given in
Fig.\ref{fig9js}:

 The reduced pion multiplicity $<n_\pi> / A_{part}$ for
 nucleus-nucleus collisions is below  that for elementary $NN$ interactions at moderate
 energies and exceeds appropriate $NN$ values in a relativistic
 domain. According to~\cite{GG99,Gazd05}, the equality of the nuclear and elementary
 $<n_\pi> / A_{part}$ ratios at the "horn" energy is treated as an
 argument in favor of the onset of deconfinement. However, this
 proximity of nuclear and elementary reduced pion multiplicities
 starts at the Dubna Nuclotron energy and could naturally be explained by the
 role of the $\Delta$ isobar in the pion absorbtion. To clarify this or
 alternative interaction mechanisms, new experimental data for heavy ion
collisions with scanning over the Nuclotron energy range are
needed.

\item   Existing experimental data indicate a large degree of
equilibration of nuclear matter from high-energy heavy-ion
collisions at freeze-out~(see review-article \cite{BRS03}). The
 results presented in Fig.\ref{figratio130} demonstrate a high quality
description of relative hadron yields within the equilibrium
statistical model obtained by means of fitting the temperature and
baryon chemical potential.

\begin{figure}[h]
\hspace*{4mm}\includegraphics[width=7.7cm,height=8cm]{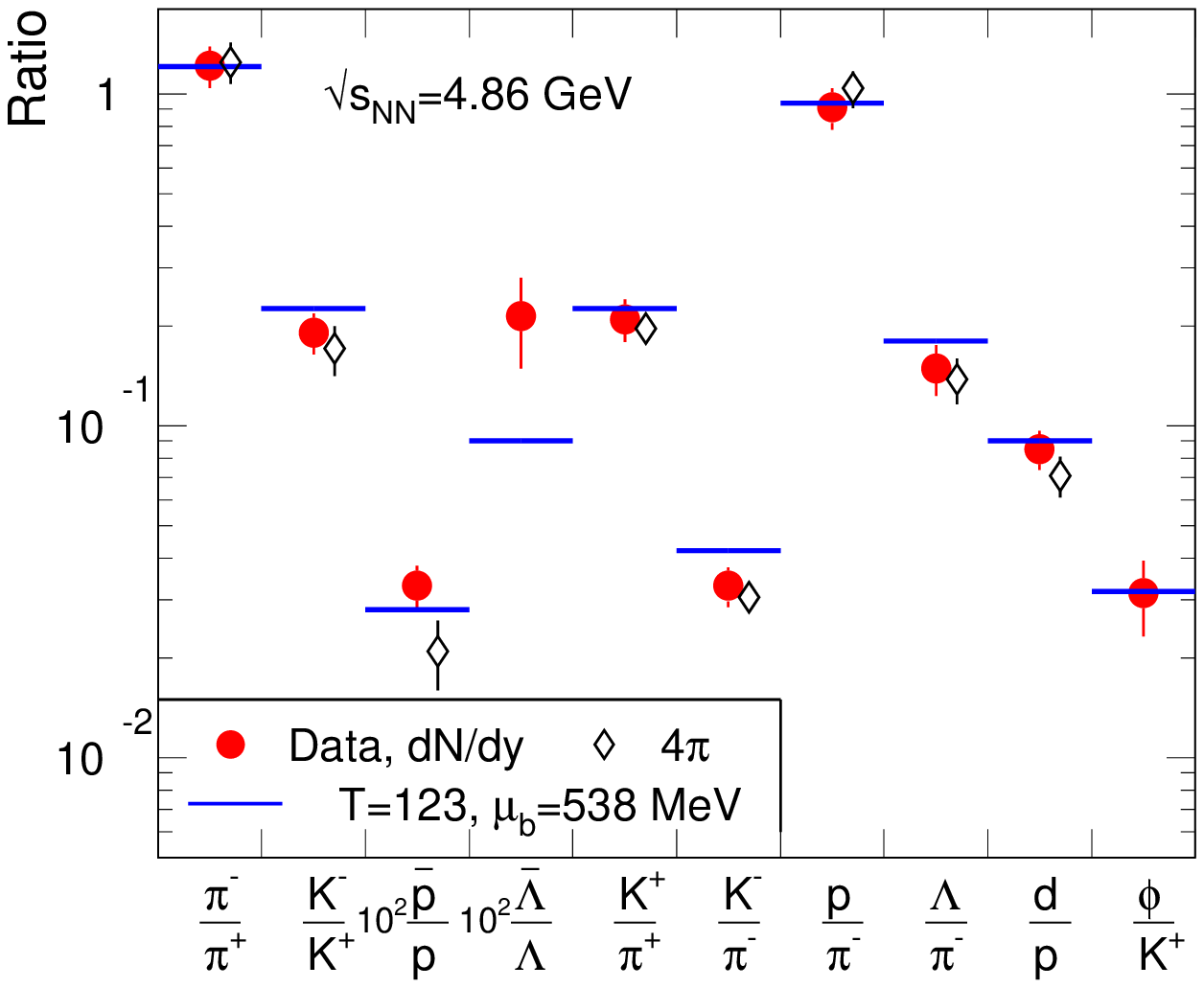}
\includegraphics[width=7.7cm,height=8cm]{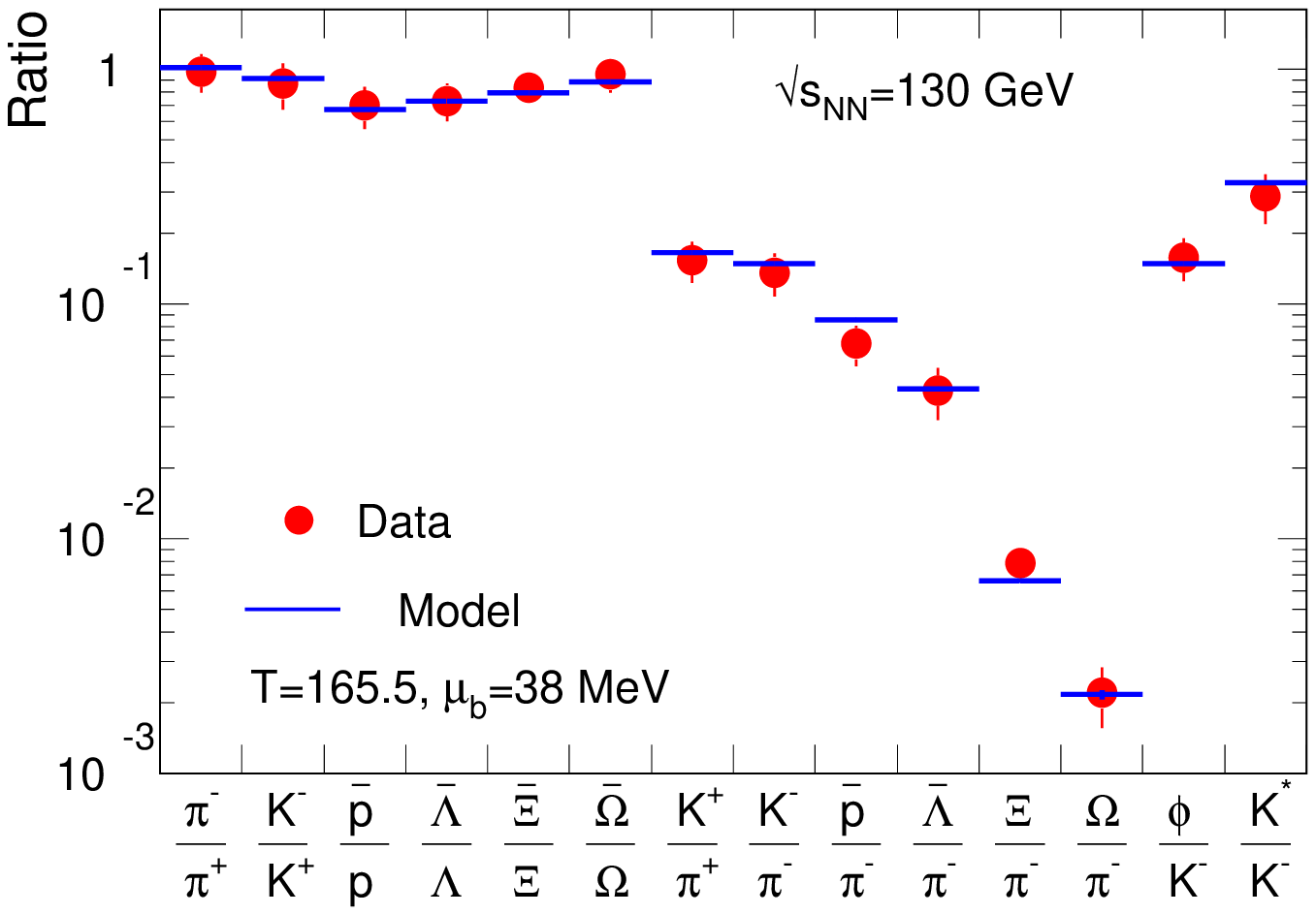}
\caption{Yield ratios with the best fit at midrapidity for the top
AGS beam energy 10.7 AGeV (left panel) and for the RHIC colliding
energy $\sqrt{s_{NN}}=$130 GeV (right panel)~\cite{ABMS05}. The
diamonds represent ratios of yields integrated over $4\pi$. Note
the scaling factor 100 for the ratios $\bar{p}/p$ and
$\bar{\Lambda}/\Lambda$ in the left figure. The extracted
freeze-out temperature and baryon chemical potential are displayed
in the figures. } \label{figratio130}
\end{figure}

 The mechanisms of equilibration are still unclear. In particular, a large difference
 between dynamical and statistical models is observed for heavy baryons (e.g. for
$\bar \Lambda$, multi-strange hyperons, $\Xi$  and $\Omega^-$  and
resonances) measured at high energies. The role of multiparticle
interactions is also obscure. The corresponding data for the
Nuclotron energies are extremely poor. Future Nuclotron
measurements
 could give a strong input for a further development of dynamical and statistical
 approaches and allow one to disentangle  these two types of models.

\item To a certain extent, the  study of fluctuations in
relativistic strongly interacting matter may help with solving the
equilibration problem mentioned above. Experimental data on
event-by-event fluctuations (e.g., fluctuations of particle
multiplicity, electric, baryon and strangeness charges)
  in nuclear collisions give a unique
possibility to test recent ideas on the origin of fluctuations in
relativistic interacting systems~\cite{Fluct1,Fluct2}. Among them
suppression of event-by-event fluctuations of electric charge
 was predicted \cite{Fluct1} as a consequence of deconfinement.
Estimates of the magnitude of the charge fluctuations indicate
that they are much smaller in a quark-gluon plasma than in a
hadron gas. Thus, naively, a decrease of the fluctuations is
expected when the collision energy crosses the threshold for the
deconfinement phase transition. However, this prediction is
derived under assumptions that  initial fluctuations survive
through hadronization and that their relaxation times in hadronic
matter are significantly longer than in the hadronic stage of the
collision \cite{Fluct1,Zar02}. The magnitude of the measured
charge fluctuations depends not only on the unit of electric
charge $q$ carried by degrees of freedom of the system (integer
charge of hadrons or fractional one of quarks), but also on
trivial effects, which may obscure the physics of interest (the
fluctuations in the event multiplicity, caused  by the variation
of the impact parameter, and changes in the mean multiplicity due
to changes of the acceptance). In our example, $\Delta\Phi_q$ is
used as a measure of charge fluctuation which is constructed from
the well established measure $\Phi_q$ of event-by-event
fluctuations, defined as:
\begin{equation}
\Phi_q = \sqrt {<Z^2>  / <N> } - \sqrt {\bar{z^2}} \nonumber
\end{equation}
 where: $z = q - \bar q, \ Z = \sum^N_{i=1} (q_i - \bar q)$. In these equations
  $N$ is the number of particles of the given event within the
acceptance, and over-line and $<...>$ denote averaging over a
single particle inclusive distribution and over all events,
respectively.

For a scenario, in which particles are correlated only by global
charge conservation (GCC), the value of $\Phi_q$ is given by
$\Phi_{q,GCC} = \sqrt{1- P}- 1$ where $P = <N_{ch}>/
<N_{ch}>_{tot}$  with $<N_{ch}>$ and $<N_{ch}>_{tot}$ being the
mean charged multiplicity in the detector acceptance and in full
phase space (excluding spectator nucleons), respectively. In order
to remove the sensitivity to GCC, the measure $\Delta \Phi_q$ is
defined as the difference
$$\Delta \Phi_q= \Phi_q -\Phi_{q,GCC}~.$$
 By construction, the value of $\Delta \Phi_q$ is
zero if the particles are correlated by global charge conservation
only. It is negative in the case of an additional correlation
between positively and negatively charged particles, and it is
positive if the positive and negative particles are
anticorrelated~\cite{Zar02}.

\begin{figure}[thb]
\includegraphics[width=12cm]{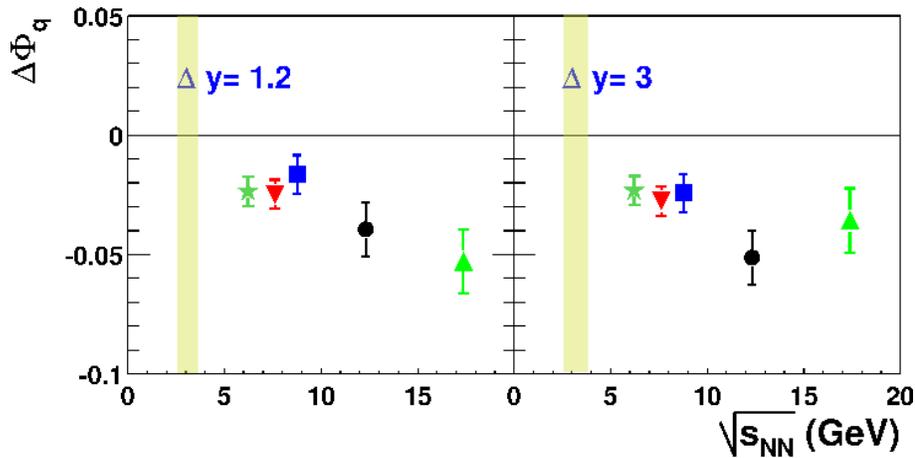}
   \caption{The colliding energy dependence of $\Delta\Phi_q$ for a small
   $\delta y=1.2$ and large $\delta y=3$ rapidity interval for
   central $Pb+Pb$ collisions~\cite{Alt04}. Shaded bands correspond to the
   Nuclotron energies. }
   \label{figj6s}
\end{figure}

As is seen from the results presented in Fig.\ref{figj6s}, the
measured values of $\Delta\Phi_q$ vary between 0 and -0.05. They
are significantly larger than those expected for QGP fluctuations
$-0.5 < \Delta\Phi_q < -0.15$.  A weak decrease of $\Delta\Phi_q$
with increasing energy is suggested by the presented data.

 Measurements require tracking detectors of  large
 acceptance and precise control of collision
centrality on event-by-event basis. Up to now only results on very
limited acceptance at high energies are available, thus  new
measurements at the Nuclotron energy are of particular importance.
From the experimental point of view the Nuclotron energy range
seems to be ideal for these measurements. This is because moderate
particle multiplicity and their relatively broad angular
distribution simplify an efficient detection of all produced
charged particles.

\item An investigation of the narrow Hanbury-Brown--Twiss
correlations~\cite{HBT} joins this problem. Momentum correlations
between mesons have been widely used to study the space-time
extent of the emitting source in nucleus-nucleus collisions. For
identical bosons the symmetry requirement of the Bose–Einstein
statistics results in an enhanced production of bosons with a
small momentum difference.   Kopylov and Podgoretsky~\cite{GKP71}
first noticed the deep analogy of this effect with Hanbury-Brown
and Twiss  space–time correlations of the classical
electromagnetic fields used in astronomy for interferometric
measurement of star angular radii and developed the basic methods
of momentum interferometry in particle and nuclear collisions.
This technique represents an important tool for investigating the
underlying reaction mechanism as far as it influences the
decoupling configuration, including an unusually long decoupling
time suggested as one of the signatures of quark-gluon plasma
formation~\cite{P86}.

The correlation function $C_2(p_1,p_2)$ is defined as the ratio of
the probability of observing particles with four-momenta $p_1$ and
$p_2$ simultaneously in one event divided by the probability for
pairs of independent particles with the same single particle phase
space distribution. In Gaussian approximation with the
Bertsch-Pratt parameterization~\cite{P86}  the correlation
function may be reduced to
$$C_2(q_{out},q_{side}, q_{long}) = 1+\lambda \ \exp{[- q^2_{out}R^2_{out} -q^2_{side}R^2_{side} -
q^2_{long}R^2_{long} ]},$$
 where $q_{long}$ is the longitudinal momentum component,
and ($q_{out}, q_{side}$) are the components of the transverse
3-momentum difference in an outward direction and perpendicular to
it. The parameter $\lambda$ characterizes the degree of
incoherence.
\begin{figure}[h]
\hspace*{2mm}\includegraphics[width=9.cm]{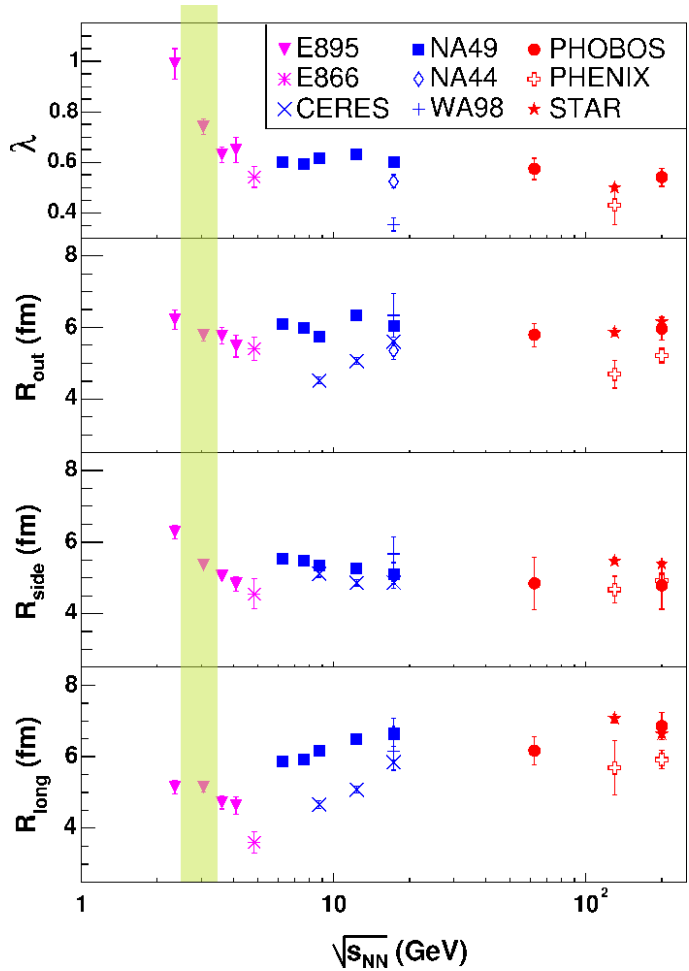}
\caption{Bertsch-Pratt parameters as a  function of colliding
energy for $\pi^-\pi^-$ pairs from central Pb+Pb/Au+Au
collisions~\cite{HBT-PHOBOS}. The presented data are near
midrapidity and for comparable transverse momentum bins from nine
different experiments. Systematic errors are not shown. }
\label{figj12}
\end{figure}

 The parameters of the HBT analysis are presented in Fig.\ref{figj12} as
a function of the colliding energy.  Although there is some
disagreement among experiments, the data do not appear to exhibit
any sharp discontinuities but rather smoothly vary as a function
of the collision energy. The detailed predictions of Gyulassy and
Rischke~\cite{RG96} about a slowly burning quark-qluon plasma have
obtained a lot of experimental attention. Namely, it was expected
that the HBT radius in the out direction will exceed the HBT
radius in the side direction, predicting $R_{out}/R_{side}
> 1$ values reaching up to 4 or even 20 in case of an ideal first
order  phase transition\cite{RG96}.  In contrast, as is seen from
Fig.\ref{figj12}, $R_{out}\approx R_{side} \approx R_{long}$  in
the RHIC energy region.

The correlation function parameters may be related to physical
ones characterizing the system size, lifetime, freeze-out
duration, and expansion time. In particular, based on a
hydrodynamic expansion model it was shown that $R_{long} = \tau_f
\sqrt{ T \over m_T}$~\cite{MS88}, where $\tau_f$ and $T$ are the
freeze-out time and temperature of the particles, respectively.

One should emphasize that all these investigations suppose that
centrality of heavy-ion collisions is under control and centrality
scanning of the characteristics under discussion is an
indispensable  condition.

\end{itemize}

Measurements of these quantities at the Nuclotron energies should
be considered as a necessary continuation of global efforts to
establish the energy dependence of properties of hadron production
and search for  signals of a phase change in nuclear collisions.

In this connection the following theoretical and experimental
studies at JINR are considered as perspective~\cite{SSSTZ05}:
\begin{enumerate}
\itemsep=-2mm \item[1)] research into the hadron properties in hot
and/or dense baryonic matter. A spectral function change is
expected, first of all for the $\sigma$-meson as a chiral partner
of pions, which characterizes a degree of chiral symmetry
violation. The rare specific channels of $\rho$-meson decays are
also quite attractive.
\end{enumerate}

\vspace*{-.3cm} \noindent
 Solving these issues assumes a proper
understanding of reaction mechanisms of high-energy colliding
ions, knowledge of properties of strongly interacting QCD matter
and its equation of state. In this respect, more general
researches are in order:

\vspace*{-.1cm}

\begin{enumerate}
\itemsep=-2mm \item[2)] analyzing multiparticle hadron
interactions, targeted to the development of a new statistical
treatment as well as codes for space-time evolution of heavy
nuclei collisions at high energies. Particular attention should be
paid to signals of a new phase formation during this evolution;

\item[3)]  studying the system size, lifetime, freeze-out
duration, expansion time in the HBT analysis, scanning in atomic
number and energy;

\item[4)]  analyzing the energy and centrality dependencies of the
pion, hadron resonance and strange particle multiplicities, and
the ratio of their yields, together with the transverse momentum,
including $K^-, K^*-$ and $\phi$-meson spectra as well as
manifestation of baryon repulsion effects on hadron abundances;

\item[5)] studying dileptons (electron and muon pairs) production
to see in-medium modification of hadron properties at high baryon
densities;

\item[6)] studying angular correlations in the transverse plane
as well as radial, directed and elliptic flows;

\item[7)]  analyzing fluctuations of multiplicities, electric
charge, and transverse momenta for secondary particles (their
energy dependencies could give information on the phase transition
range);

\item[8)] analyzing nuclear fragments characteristics versus
centrality, universality of nuclear fragmentation;

\item[9)]  energy and atomic number scanning for all
characteristics of central heavy nuclei collisions (this might
allow one to obtain information on the equation of state of
strongly interacting QCD matter in the transition region),
difference between central collisions of light nuclei and
peripheral heavy ion collisions.
\end{enumerate}

The JINR Nuclotron has a possibility to accelerate heavy ions (up
to $A>$200) to the maximal energy of 5 AGeV in about a year. This
gives a chance to address experimentally many recent problems
within the next several years before the FAIR GSI accelerator
comes into operation. The proposed research program at the
Nuclotron may be considered as a pilot study preparing for
subsequent detailed investigations at SIS-100/300~\cite{GSI300}
and as an integral part of the world scientific cooperation to
study the energy dependence of hadron production properties in
nuclear collisions.

\vspace*{3mm}
 {\bf Acknowledgements}
We greatly appreciate many useful and valuable discussions with M.
Gazdzicki, M. Gorenstein, H. Gutbrod T. Hatsuda, T. Kunihiro, H.
Satz, and H. Str\"obele. We are thankful to V.V. Skokov for help
in preparing the manuscript.
This work was supported in part by RFBR Grant N 05-02-17695.


\begin{thebibliography}{99}

\bibitem{GSI300}Proposal for an International Accelerator Facility
for Research with Heavy Ions and Antiprotons,
http://www.gsi.de/documents/DOC-2004-Mar-196-2.pdf~.
%
\bibitem{Gprog} M. Gazdzicki, arXiv:nucl-ex/0512034.
%
\bibitem{IRT05} Y.B.~Ivanov, V.N.~Russkikh and V.D.~Toneev,
nucl-th/0503088.
%
\bibitem{ZS} E.V.~Shuryak and I.~Zahed, hep-ph/0307267.
%
\bibitem{V04}D.N.~Voskresensky,
Nucl. Phys. A {\bf 744} (2004)  378 [arXiv:hep-ph/0402020];
 G.E.~Brown, Ch,-H.~Lee, and M.~Rho,
A new state of matter at high temperature as "sticky molasses",
arXiv:hep-ph/0402207.
%
\bibitem{I05} Yu.~Ivanov, Multi-fluid
hydrodynamics, Talk at the CBM Collaboration Meeting "FAIR, The
physics of compressed baryonic matter", December 15-16, 2005, GSI,
Darmstadt, http://www.gsi.de/documents/DOC-2005-Dec-87-112-1.pdf;
V.N.~Russkikh, privite communication.
%
\bibitem{Gazd05} M.~Gazdzicki, arXiv:nucl-ex/0507017; C.~Blume (NA49 Collaboration),
[arXiv:hep-ph/0505137].
%
\bibitem{Klay03} J.L.~Klay
et al. Phys. Rev. C {\bf 68}, 054905 (2003)
[arXiv:nucl-ex/0306033].
%
\bibitem{Holzman02} B.~Holzman et al.
Nucl. Phys. {\bf A698} 643 (2002) [arXiv:nucl-ex/0103015].
%
\bibitem{Wol05} G.~Wolschin,
arXiv:hep-ph/0509108.
%
\bibitem{TK94}T.~Hatsuda and T.~Kunihiro, Phys. Rep. {\bf 247} (1994)
221.
%
\bibitem{BR96} G.E.~Brown and M.~Rho, Phys. Rep. {\bf 269}, (1996) 333.
%
\bibitem{CERES40} D.~Adamova et al.,  CERES/NA45 Collaboration, arXiv:nucl-ex/0209024.
%
\bibitem{RW00} R.~Rapp and J.~Wambach, Adv. Nucl. Phys. {\bf 25},
(2000) 1.
%
 \bibitem{NA60-QM05}S.~Scomparin et al., QM 2005 Proceedings (2005);
S.~Damjanovic et al., QM 2005 Proceedings (2005)
[arXiv:nucl-ex/0510044].
%
\bibitem{ST05} V.V~Skokov and V.D.~Toneev, arXiv:nucl-th/0509085.
%
\bibitem{CHAOS} F.~Bonutti {\it et al.}, (CHAOS Collabortion), Phys. Rev. Lett.
{\bf 77}, (1996) 603; Nucl. Phys. {\bf A677} (2000) 213.
%
 \bibitem{CBall} A.~Starostin {\it et al.}, Crystal Ball Collaboration,
 Phys. Rev. Lett. {\bf 85} (2000) 5539.
 %
\bibitem{TAPS2} CBELSA/TAPS Collaboration, nucl-ex/0504010.
%
\bibitem{CH98}S.~Chiku and T.~Hatsuda, Phys. Rev. {\bf D58}, 076001 (1998) [hep-ph/9809215];
T. Hatsuda, T. Kunihiro and H.~Shimizu, Phys. Rev. Lett. {\bf 82},
(1999) 2840.
%
\bibitem{VKBRS98}  M.K.~Volkov, E.A.~Kuraev, D.~Blaschke, G.~R\"opke and S.M.~Schmidt,
Phys. Lett. {\bf B424}, (1998) 235 [hep-ph/9706350].
%
\bibitem{Weld92} H.A. Weldon, Phys. Lett. {\bf B 274} (1992) 133.
%
\bibitem{St04} R.~Stock, arXiv:nucl-ex/0405007.
%
\bibitem{Ch01} P.~Chung et al.,  arXiv:nucl-ex/0112002.
%
 \bibitem{GG99}M.~Gazdzicki and M. I.~Gorenstein, Acta Phys. Polon. {\bf B
 30}, (1999) 2705 [arXiv:hep-ph/9803462].
%
\bibitem{Gazd04} M.~Gazdzicki (for the NA49 collaboration), J. Phys. {\bf G 30} (2004), S701.
%
\bibitem{WBCS02} H.~Weber, E.L.~Bratkovskaya, W.~Cassing, and H.~Stoecker,
nucl-th/0209079.
%
\bibitem{Schn94}E.~Schnedermann and U.W.~Heinz, Phys. Rev. C {\bf 50}, 1675
(1994)[arXiv:nucl-th/9402018]; E.~Schnedermann, J.~Sollfrank,
U.W.~Heinz,. Phys. Rev. C {\bf 48}, 2462 (1993)
[arXiv:nucl-th/9307020].
%
\bibitem{vH82} L. Van Hove, Phys. Lett. {\bf B 118}, 138 (1982).
%
\bibitem{MASNN03}B.~Mohanty, Jan-e~Alam, S.~Sarkar, T.K.~Nayak,
B.K Nandi, Phys. Rev. C {\bf 68}, 021901 (2003)
[arXiv:nucl-th/0304023].
%
\bibitem{hydro04} M.~Gazdzicki, et al., Braz. J. Phys. {\bf 34},
322 (2004)[arXiv:hep-ph/0309192].
%
\bibitem{Bl05} C.~Blume. J. Phys. G: Nucl. Part. Phys. {\bf 31}, S57
(2005).
%
\bibitem{SS99} P.~Senger and H.~Str\"obele, J. Phys. G: Nucl. Part.
Phys. {\bf 25} (1999) R59.
%
\bibitem{BRS03}P. Braun-Munzinger, K. Redlich and J. Stachel,
in {\em Quark Gluon Plasma 3} eds. R.C.~Hwa  and X.N.~Wang, World
Scientific, Singapore, p.491 [nucl-th/0304013].
%
\bibitem{ABMS05} A. Andronic, P. Braun-Munzinger, J.
Stachel, nucl-th/0511071.

%
\bibitem{Fluct1} S.~Jeon and V.~Koch, Phys. Rev. Lett. {\bf 85}, 2076 (2000)
[arXiv:hep-ph/0003168]. M.~Asakawa, U.W.~Heinz and B.~Muller,
Phys. Rev. Lett. {\bf 85}, 2072 (2000) [arXiv:hep-ph/0003169].
E.V.~Shuryak and M.A. ~Stephanov, Phys. Rev. C {\bf 63}, 064903
(2001) [arXiv:hep-ph/0010100].
%
\bibitem{Fluct2}  V.V.~Begun, M.I.~Gorenstein,
A.P.~Kostyk and O.S.~Zozulya, Phys.   Rev. {\bf C 71}, 054904
(2005); A.~Keranen, F.~Becattini V.V.~Begun, M.I.~Gorenstein and
O.S.~Zozulya, J. Phys. {\bf G 31}, S1095 (2005); V.V.~Begun,
M.~Gazdzicki, M.I.~Gorenstein and O.S.~Zozulya, Phys. Rev. {\bf C
71}, (2005) 054904.
%
\bibitem{Zar02} J.~Zaranek, Phys. Rev. C {\bf 66}, 024905 (2002)
[arXiv:hep-ph/0111228].
%
\bibitem{Alt04} C.~Alt et al., Phys. Rev. C {\bf 70}, 064903 (2004)
[arXiv:nucl-ex/0406013].
%
\bibitem{HBT} U.A.~Wiedemann and U.~Heinz, Phys. Rept. {\bf 319}, 145
(1999); T.~Csorgo, Heavy Ion Phys. {\bf 15}, 1 (2002);
R.~Lednicky, Phys. At. Nucl. {\bf 67}, 72 (2004).
 %
\bibitem{GKP71}
G.I. Kopylov and M.I. Podgoretsky, Sov. J. Nucl. Phys. 15 (1972)
219 ; ibid. 18 (1973) 336; Sov. Phys. JEPT 42 (1975); G.I.
Kopylov, Phys. Lett. 50 (1974) 472; M.I. Podgoretsky, Sov. J.
Part. Nucl. 20 (1989) 266.
%
\bibitem{P86} S. Pratt, Phys. Rev. {\bf D33}, 1314 (1986); G. Bertsch et al.,
Phys. Rev. {\bf D37}, 1202 (1988).
%
\bibitem{HBT-PHOBOS}  B.B.~Back, et al., PHOBOS Collaboration,
 arXiv:nucl-ex/0409001.
%
\bibitem{RG96}D. H. Rischke and M. Gyulassy, Nucl. Phys. {\bf A 597}, 701 (1996)
[arXiv:nucl-th/9509040]; ibid. {\bf A 608}, 479 (1996)
[arXiv:nucl-th/9606039].
%
\bibitem{MS88} A. N. Makhlin and Y. M. Sinyukov, Z. Phys. {\bf C39} 69
(1988).
%
\bibitem{SSSTZ05} A.N.~Sissakian, A.S.~Sorin, M.K.~Suleymanov, V.D.~Toneev, and G.M.~Zinovjev,
arXiv:nucl-ex/0511018.



\end{thebibliography}
\end{document}